\documentclass[letter,traditabstract,longauth]{aa} 
\usepackage{graphicx}
\usepackage{txfonts}
\begin{document}
   \title{HerMES: Halo Occupation Number and Bias Properties of Dusty Galaxies from Angular Clustering Measurements\thanks{Herschel is an ESA space observatory with science instruments provided by European-led Principal Investigator consortia and with important participation from NASA.}}

\author{Asantha Cooray\inst{1,2} 
\and A.~Amblard\inst{1}
\and L.~Wang\inst{3}
\and B.~Altieri\inst{4}
\and V.~Arumugam\inst{5}
\and R.~Auld\inst{6}
\and H.~Aussel\inst{7}
\and T.~Babbedge\inst{8}
\and A.~Blain\inst{2}
\and J.~Bock\inst{2,9}
\and A.~Boselli\inst{10}
\and V.~Buat\inst{10}
\and D.~Burgarella\inst{10}
\and N.~Castro-Rodr{\'\i}guez\inst{11,12}
\and A.~Cava\inst{11,12}
\and P.~Chanial\inst{8}
\and D.L.~Clements\inst{8}
\and A.~Conley\inst{13}
\and L.~Conversi\inst{4}
\and C.D.~Dowell\inst{2,9}
\and E.~Dwek\inst{14}
\and S.~Eales\inst{6}
\and D.~Elbaz\inst{7}
\and D.~Farrah\inst{3}
\and M.~Fox\inst{8}
\and A.~Franceschini\inst{15}
\and W.~Gear\inst{6}
\and J.~Glenn\inst{13}
\and M.~Griffin\inst{6}
\and M.~Halpern\inst{16}
\and E.~Hatziminaoglou\inst{17}
\and E.~Ibar\inst{18}
\and K.~Isaak\inst{6}
\and R.J.~Ivison\inst{18,5}
\and A.A.~Khostovan\inst{1}
\and G.~Lagache\inst{19}
\and L.~Levenson\inst{2,9}
\and N.~Lu\inst{2,20}
\and S.~Madden\inst{7}
\and B.~Maffei\inst{21}
\and G.~Mainetti\inst{15}
\and L.~Marchetti\inst{15}
\and G.~Marsden\inst{16}
\and K.~Mitchell-Wynne\inst{1}
\and A.M.J.~Mortier\inst{8}
\and H.T.~Nguyen\inst{9,2}
\and B.~O'Halloran\inst{8}
\and S.J.~Oliver\inst{3}
\and A.~Omont\inst{22}
\and M.J.~Page\inst{23}
\and P.~Panuzzo\inst{7}
\and A.~Papageorgiou\inst{6}
\and C.P.~Pearson\inst{24,25}
\and I.~P{\'e}rez-Fournon\inst{11,12}
\and M.~Pohlen\inst{6}
\and J.I.~Rawlings\inst{23}
\and G.~Raymond\inst{6}
\and D.~Rigopoulou\inst{24,26}
\and D.~Rizzo\inst{8}
\and I.G.~Roseboom\inst{3}
\and M.~Rowan-Robinson\inst{8}
\and M.~S\'anchez Portal\inst{4}
\and B.~Schulz\inst{2,20}
\and Douglas~Scott\inst{16}
\and P.~Serra \inst{1}
\and N.~Seymour\inst{23}
\and D.L.~Shupe\inst{2,20}
\and A.J.~Smith\inst{3}
\and J.A.~Stevens\inst{27}
\and M.~Symeonidis\inst{23}
\and M.~Trichas\inst{8}
\and K.E.~Tugwell\inst{23}
\and M.~Vaccari\inst{15}
\and I.~Valtchanov\inst{4}
\and J.D.~Vieira\inst{2}
\and L.~Vigroux\inst{22}
\and R.~Ward\inst{3}
\and G.~Wright\inst{18}
\and C.K.~Xu\inst{2,20}
\and M.~Zemcov\inst{2,9}}

\institute{Dept. of Physics \& Astronomy, University of California, Irvine, CA 92697, USA\\
 \email{acooray@uci.edu}
\and California Institute of Technology, 1200 E. California Blvd., Pasadena, CA 91125, USA
\and Astronomy Centre, Dept. of Physics \& Astronomy, University of Sussex, Brighton BN1 9QH, UK
\and Herschel Science Centre, European Space Astronomy Centre, Villanueva de la Ca\~nada, 28691 Madrid, Spain
\and Institute for Astronomy, University of Edinburgh, Royal Observatory, Blackford Hill, Edinburgh EH9 3HJ, UK
\and Cardiff School of Physics and Astronomy, Cardiff University, Queens Buildings, The Parade, Cardiff CF24 3AA, UK
\and Laboratoire AIM-Paris-Saclay, CEA/DSM/Irfu - CNRS - Universit\'e Paris Diderot, CE-Saclay, pt courrier 131, F-91191 Gif-sur-Yvette, France
\and Astrophysics Group, Imperial College London, Blackett Laboratory, Prince Consort Road, London SW7 2AZ, UK
\and Jet Propulsion Laboratory, 4800 Oak Grove Drive, Pasadena, CA 91109, USA
\and Laboratoire d'Astrophysique de Marseille, OAMP, Universit\'e Aix-marseille, CNRS, 38 rue Fr\'ed\'eric Joliot-Curie, 13388 Marseille cedex 13, France
\and Instituto de Astrof{\'\i}sica de Canarias (IAC), E-38200 La Laguna, Tenerife, Spain
\and Departamento de Astrof{\'\i}sica, Universidad de La Laguna (ULL), E-38205 La Laguna, Tenerife, Spain
\and Dept. of Astrophysical and Planetary Sciences, CASA 389-UCB, University of Colorado, Boulder, CO 80309, USA
\and Observational  Cosmology Lab, Code 665, NASA Goddard Space Flight  Center, Greenbelt, MD 20771, USA
\and Dipartimento di Astronomia, Universit\`{a} di Padova, vicolo Osservatorio, 3, 35122 Padova, Italy
\and Department of Physics \& Astronomy, University of British Columbia, 6224 Agricultural Road, Vancouver, BC V6T~1Z1, Canada
\and ESO, Karl-Schwarzschild-Str. 2, 85748 Garching bei M\"unchen, Germany
\and UK Astronomy Technology Centre, Royal Observatory, Blackford Hill, Edinburgh EH9 3HJ, UK
\and Institut d'Astrophysique Spatiale (IAS), b\^atiment 121, Universit\'e Paris-Sud 11 and CNRS (UMR 8617), 91405 Orsay, France
\and Infrared Processing and Analysis Center, MS 100-22, California Institute of Technology, JPL, Pasadena, CA 91125, USA
\and School of Physics and Astronomy, The University of Manchester, Alan Turing Building, Oxford Road, Manchester M13 9PL, UK
\and Institut d'Astrophysique de Paris, UMR 7095, CNRS, UPMC Univ. Paris 06, 98bis boulevard Arago, F-75014 Paris, France
\and Mullard Space Science Laboratory, University College London, Holmbury St. Mary, Dorking, Surrey RH5 6NT, UK
\and Space Science \& Technology Department, Rutherford Appleton Laboratory, Chilton, Didcot, Oxfordshire OX11 0QX, UK
\and Institute for Space Imaging Science, University of Lethbridge, Lethbridge, Alberta, T1K 3M4, Canada
\and Astrophysics, Oxford University, Keble Road, Oxford OX1 3RH, UK
\and Centre for Astrophysics Research, University of Hertfordshire, College Lane, Hatfield, Hertfordshire AL10 9AB, UK}


 
  \abstract
   {We measure the angular correlation function, $w(\theta)$, from 0.5 to 30 arcminutes 
of detected sources in two wide fields of the Herschel Multi-tiered Extragalactic Survey (HerMES). 
Our measurements are consistent with the expected clustering shape from a population of sources that trace the
dark matter density field, including non-linear clustering at arcminute angular scales arising from multiple
sources that occupy the same dark matter halos.  By making use of the halo model to connect the spatial clustering of 
sources to the dark matter halo distribution, 
we estimate source bias and halo occupation number for dusty sub-mm galaxies at $z \sim 2$.
We find that sub-mm galaxies with $250\,\mu$m flux densities above 30 mJy reside in dark matter halos with mass above
$(5\pm4)\times10^{12}$M$_{\sun}$, while $(14\pm8)$\% of such sources appear as satellites in more massive halos.
}


\titlerunning{Angular Correlation Function of sources in HerMES}

   \maketitle
%

\section{Introduction}

The Herschel Multi-tiered Extragalactic Survey (HerMES\footnote{hermes.sussex.ac.uk}; Oliver et al. 2010) is the largest project 
being undertaken by {\it Herschel} (Pilbratt et al. 2010).
It surveys a large set of well-known extra-galactic fields (totaling 70 deg$^2$) at various depths,
primarily using the Spectral and Photometric Imaging Receiver (SPIRE) instrument (Griffin et al. 2010).  

In this {\it Letter}, we focus on the clustering of these sources from 0.5 to 30 arcminute angular scales 
by making use of source catalogues in the two widest HerMES  fields,
Lockman-SWIRE and the {\it Spitzer} First Look Survey (FLS) observed during the Science Demonstration Phase (Oliver et al. 2010). 
Previous studies on the spatial correlations  of sub-mm galaxies
was limited to at most 100 sources, leading either to a limit on the clustering amplitude (Blain et al. 2004)
or a marginal detection (Scott et al. 2006).  While the {\it BLAST} source catalog was not used for a measurement of the angular
correlation function, clustered fluctuations were detected in a power spectrum analysis of all three bands (Viero et al. 2009).

In the Lockman-SWIRE field we have detected 8154, 4899, and 1680 sources with flux densities above 30 mJy at 250, 350, and $500\,\mu$m, respectively,
in an area of $218'\times218'$ (Oliver et al. 2010). These counts are supplemented by 3592, 2207, and 1016  sources detected in the FLS field over
an area of $155'\times135'$, again down to the
same flux density in each of the three bands. These numbers allow clustering estimates at the same precision level as the
first-generation of clustering studies at shorter IR wavelengths with source samples from
{\it Spitzer} data (e.g.,  Farrah et al. 2006; Magliocchetti et al. 2007; Waddington et al. 2007; Brodwin et al. 2008).
Instead of simple power-law models, the correlation functions of HerMES sources have high enough 
signal-to-noise ratios that we  are also able to constrain parameters of a halo model (e.g. Cooray \& Sheth 2002). 


\section{$w(\theta)$ measurement}

The angular correlation function, $w(\theta)$, is a measure of the probability above Poisson fluctuations of finding two galaxies with a separation $\theta$,
$Pd\Omega_1d\Omega_2 =N[1+w(\theta)]d\Omega_1d\Omega_2$, 
where $N$ is the surface density of galaxies and $d\Omega_i$ are solid angles for each galaxy,
 corresponding to angle $\theta$.
The angular correlation function is of great interest in  cosmology as sources are expected to 
trace the underlying  dark matter distribution and the clustering of sources can be related to that of the dark matter halos.

For clustering measurements we make use of the HerMES source catalogues (Oliver et al. 2010),
based on maps made with calibrated timelines with optimized internal astrometry.
The maps were produced using the standard SPIRE pipeline after undergoing 
calibration and other reduction procedures (Swinyard et al. 2010). 
In the FLS and Lockman-SWIRE fields, a small number of individual scans have been removed due to 
artifacts arising from the temperature drift correction.  
Catalogues were generated using the SUSSEXtractor source extractor in {\it HIPE} version 3.0, using a Gaussian PSF with 
FWHM of 18.15$''$, 25.15$''$, and 36.3$''$ at 250, 350, and 500 $\mu$m, respectively.
Two independent maps were also produced by dividing the data in time. The sources that do not
appear in both sub-maps are flagged as spurious sources and removed from the catalogues. The overall 
astrometry has been adjusted by comparing with radio positions. We also apply a 
Wiener filter optimized for unresolved point sources to the maps in the wide fields to remove contamination from cirrus
and then we correct source fluxes through simulations. The two fields chosen for this study are largely free from cirrus
and the clustering measured with two source samples with and without the Wiener filter applied agree with each other
within the errors. Moreover, the source sample used here is restricted to sources detected with
an overall significance higher than $5\sigma$, including confusion noise.

Once the catalogues are generated, we use the Landy-Szalay estimator to measure the correlation function with
$\hat{w}(\theta) = [DD(\theta) - 2DR(\theta) + RR(\theta)]/RR(\theta)$,
where  $DD(\theta)$ is the number of unique pairs of real sources with separation $\theta$,
$DR(\theta)$ is the number of unique pairs  between the real catalogue and a mock sample of sources
with random positions, and $RR(\theta)$ is the number of unique pairs in the random source 
catalogues (Landy \& Szalay 1993).  We employ $10^3$ random mocks with $10^5$ sources in each, 
a larger number of sources than in real data to reduce shot-noise in the random pair counts.
Since the measured $w(\theta)$  does not automatically satisfy the integral constraint (Infante 1994),
we  estimate the constant necessary to correct $w(\theta)$
following Adelberger et al. (2005).   With a size of 10 deg$^2$ for each of our two fields, 
we find a required value of $\sim (2 \pm 0.2)\times10^{-3}$ between 30$'$ and 40$'$. 
This correction is small compared to $w(\theta \sim 30')\gtrsim10^{-2}$.

   \begin{figure}
   \centering
\includegraphics[width=8cm]{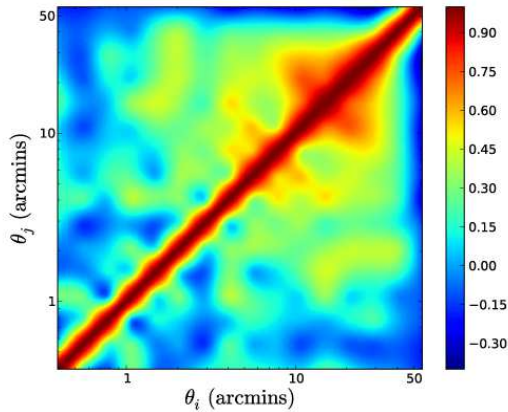}
   \caption{Correlation matrix of the angular correlation function, $C_{ij}/\sqrt{C_{ii}C_{jj}}$, where $C_{ij}\equiv\langle w(\theta_i) w(\theta_j)\rangle-\langle w(\theta_i) \rangle \langle w(\theta_j)\rangle$ at two different angular scales $\theta_i$ and $\theta_j$.
Here we show an example case at 350 $\mu$m for sources in the Lockman-SWIRE field with flux densities greater than 30 mJy.}
              \label{cij}%
    \end{figure}

   \begin{figure}
   \centering
\includegraphics[width=7.2cm]{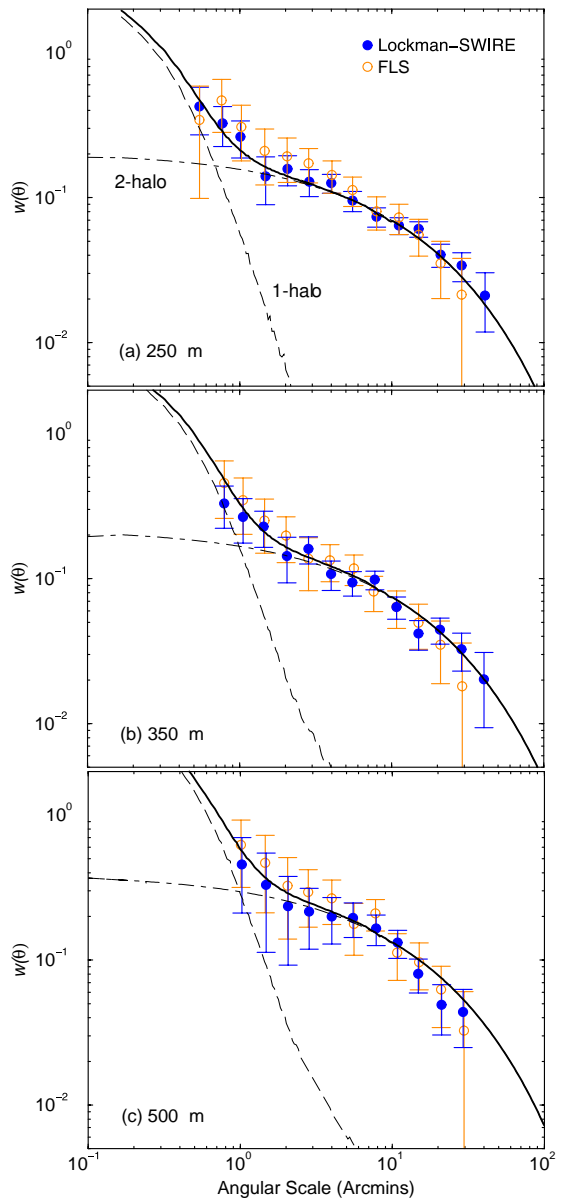}
\caption{Angular correlation function of SPIRE sources in Lockman-SWIRE and FLS with flux densities above 30 mJy: (a) 250 $\mu$m; (b) 350 $\mu$m; and
(c) 500 $\mu$m. The lines are illustrative halo models consistent with best-fit results
for the occupation number (see Table~1), with the dot-dashed lines showing the 2-halo term traced by linear clustering and the long-dashed lines showing the 1-halo 
term coming from multiple sources within the same halo. The solid lines show the total correlation function from our models.}
              \label{wtheta}%
    \end{figure}

   \begin{figure}
   \centering
\includegraphics[width=7.0cm]{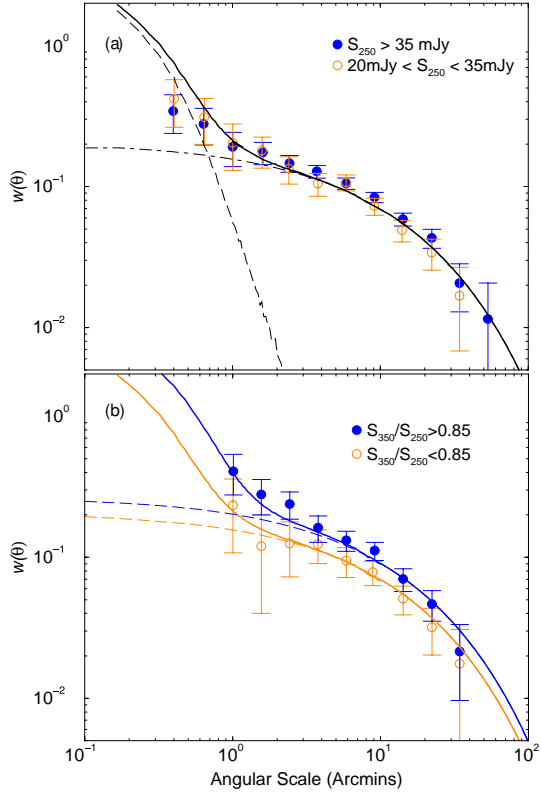}
   \caption{(a)  Angular clustering of 250 $\mu$m  sources in Lockman-SWIRE field divided into flux densities between 20 and 35 mJy and above 35 mJy.
(b)   Angular clustering for the combined 250 $\mu$m and 350 $\mu$m sample again in the Lockman-SWIRE field
with $S_{350}/S_{250}<0.85$ or $>0.85$.}
              \label{fluxcolor}%
    \end{figure}

   \begin{figure}
   \centering
\includegraphics[angle=-90,width=7cm]{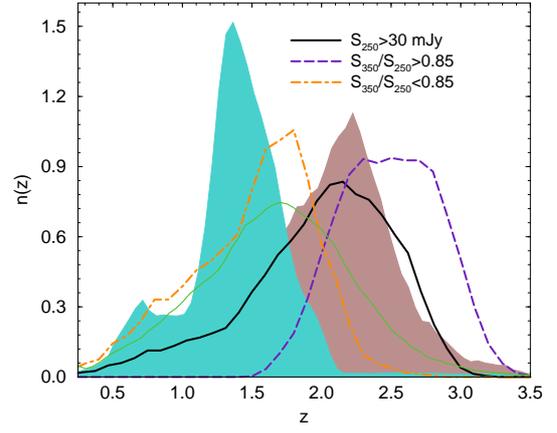}
   \caption{Approximate redshift distribution of  sources in the Lockman-SWIRE field with $S_{250} > 30$ mJy (thick solid line in black) 
and for the two cases based involving colour cuts with $S_{350}/S_{250}>0.85$ (magenta dashed line) and $S_{350}/S_{250}<0.85$ (orange
dot-dashed line) (see text for details).  
The thin solid green line and the two shaded regions in the background show example predictions for the $S_{250} > 30$ mJy
sample  and the two colour cuts, respectively, using models from
Le Borgne  et al. (2009; thin green line) and Valiante et al. (2009; shaded regions). 
}
              \label{pz}%
    \end{figure}

We calculate the covariance matrix $C_{ij}$ of the correlation function involving measurements at two different angular scales $\theta_i$ and $\theta_j$, 
using a bootstrap method similar to the one employed by Scranton et al. (2002) for measurements of angular clustering in SDSS DR1.
We also calculate the same covariance analytically following the prescription of Eisenstein \& Zaldarriaga (2001) and
find 10\% to 20\% smaller off-diagonal correlations than obtained by bootstrapping the data. 
We compute the usual Poisson errors by taking the square root of the number of pairs,
finding that the bootstrap variances are a factor of 1.5 to 2 larger and are a better representation of errors than the Poisson errors.
As an example, we show the correlation matrix of $S_{350}>30$mJy sources in the Lockman-SWIRE field in Fig.~1, where the
correlation matrix is defined as $C_{ij}/\sqrt{C_{ii}C_{jj}}$.

Since we measure $w(\theta)$ directly in the real catalogues, it is likely to be affected by a variety of effects including
source blending, flux boosting, and map-making artifacts, among others. To account for all these effects, we compute the transfer function
necessary to correct $w(\theta)$ through a set of simulations with number counts and source clustering consistent with data
and in a field similar to Lockman-SWIRE.
These input maps are then processed by the SPIRE Instrument Simulator (Sibthorpe et al. 2009) for an observational 
program exactly the same as HerMES observations of Lockman-SWIRE. The output time-ordered data from the simulator are processed as identically as
the real data. The transfer function is
defined as the ratio of the correlation function in output catalogues to that of the known input used to generate the simulations.
The transfer function is generally consistent with unity at angular scales of 10$'$ or more, but varies 25\% (at 250 $\mu$m) to 50\% (500 $\mu$m)
at angular scales of 3$'$.  We correct the measured $w(\theta)$ in data based on the average of the ratio between input and output $w(\theta)$ 
computed  from about 10 simulations. Due to the finite set of simulations we used the correction is only known to the level of
20\% between 1$'$ and 3$'$ angular scales. In comparison, at the same 1$'$ scale,  $w(\theta)$ is uncertain at the 30\% level at 
250 $\mu$m and the 50\%  level at 500 $\mu$m. Thus the uncertainty introduced by the error in the transfer function
is insignificant compared to the overall uncertainties of the measured correlation function in data.

   \begin{table*}
     \begin{center}
      \caption[]{Halo model results using the Lockman-SWIRE $w(\theta)$}
         \label{Halo}
         \begin{tabular}{ccccccccc}
            \hline
            \noalign{\smallskip}
        Band & Flux density & $N_{\rm gal}$      &  $\langle z \rangle$ & $\log [M_{\rm min}/M_{\sun}]$ & $\log [M_{\rm sat}/M_{\sun}]$ & $\alpha_s$ & $\langle b \rangle_z$ & $f_s$ \\
            \noalign{\smallskip}
            \hline
            \noalign{\smallskip}
             250$\mu$m & $S \gtrsim 30$mJy & 8154 & $2.1^{+0.4}_{-0.7}$ & $12.6^{+0.3}_{-0.6}$ & $13.1^{+0.3}_{-0.5}$ & $1.3\pm0.4$ & $2.9 \pm 0.4$& $0.14 \pm 0.08$ \\
             350$\mu$m & $S \gtrsim 30$mJy & 4899 &  $2.3^{+0.4}_{-0.7}$ & $12.9^{+0.4}_{-0.6}$ & $>13.1$ & $<1.8$ & $3.2 \pm 0.5$ & $<0.20$\\            
             500$\mu$m & $S \gtrsim 30$mJy & 1680 & $2.6^{+0.3}_{-0.7}$ & $13.5^{+0.3}_{-1.0}$ & $>13.5$ & $<1.6$ & $3.6 \pm 0.8$& $<0.24$ \\
             Combined & $S_{\rm 350}/S_{\rm 250} \gtrsim 0.85$ & 3333 & $2.5 \pm 0.4$  & $13.4^{+0.2}_{-0.3}$ & $>13.4$ & $<1.8$ & $3.4 \pm 0.6$& $<0.19$\\
            Combined & $S_{\rm 350}/S_{\rm 250} \lesssim 0.85$ & 3194 & $1.7^{+0.5}_{-0.6}$ & $12.8^{+0.3}_{-0.5}$ &  $>12.9$ & $<1.9$  & $2.6 \pm 0.6$ & $<0.26$\\
  \noalign{\smallskip}
            \hline
         \end{tabular}
\end{center}
See text below eq. (2) for  definitions of $M_{\rm min}$, $M_{\rm sat}$, $\alpha_s$, $\langle b \rangle_z$, and $f_s$.
The redshift range is an approximate estimate based on the colour-colour diagram of the source sample through a comparison to
isothermal, modified black-body SEDs with a wide range of dust temperatures and emissivity parameters (see, Fig.~4 for an example
involving $S_{250}>30$ mJy and for the two colour cuts).
   \end{table*}

\section{Halo Modeling of Angular Clustering}

In terms of the underlying three-dimensional power spectrum  
of sources as a function of redshift, $P_{\rm s}(k,z)$, the projected angular correlation function is
\begin{equation}
w(\theta) = \int dr\, n^2(r)  \int \frac{k dk}{2\pi} P_{\rm s}(k,r) J_0(kr\theta) \, ,
\end{equation}
where $r(z)$ is the radial comoving distance, $J_0(x)$ is the zeroth order Bessel function
and $n(r)$ is the radial distribution of sources normalized to unity: $\int dr\, n(r)=1$.

To model $P_{\rm s}(k)$, we make use of the halo model with both 1- and 2-halo terms (Cooray \& Sheth 2002).
While the 2-halo term captures the large-scale clustering with the linear power spectrum scaled by the source bias,
the 1-halo term captures the non-linear clustering arising from having multiple sources within a single dark matter halo.
In this case, multiple sources within a halo are subdivided to a single source at the halo center and one or more satellites.
For the central and satellite sources, we describe halo occupation numbers as
\begin{eqnarray}
\langle N_{\rm cen}(M)\rangle&=& \frac{1}{2}\left[ 1+ erf \left(\frac{\log M - \log M_{\rm min}}{\sigma_{\log M}}\right)\right] \,, \nonumber \\
\langle N_{\rm sat}(M)\rangle &=& \frac{1}{2}\left[ 1+ erf \left(\frac{\log M - \log 2 M_{\rm min}}{\sigma_{\log M}}\right)\right] 
\left(\frac{M}{M_{\rm sat}}\right)^{\alpha_{\rm s}} \,,
\end{eqnarray}
respectively, where $erf(x)$ is the error function, $M_{\rm min}$ is the minimum halo mass above which all halos host a central galaxy, and
the scatter in the relation between galaxy halo mass and luminosity is captured by $\sigma_{\log M}$. 
We take a fixed value of 0.3 for $\sigma_{\log M}$, motivated by modeling of clustering at near-IR wavelengths.
$M_{\rm sat}$ is the mass scale at which one satellite galaxy per halo is found, in addition to the central galaxy, 
and $\alpha_{\rm s}$ is the power-law 
slope of the satellite occupation number with halo mass.
In addition to these parameters, we also compute the linear bias factor of the sources, which should be interpreted as the average bias
factor of the source sample given the redshift distribution $\langle b_z \rangle$, and the satellite fraction $f_s$,
the fraction of sources in a given sample that appear as satellites in massive dark matter halos. This is calculated by taking the ratio
of number density of satellites to the total number density of sources where the number density is calculated through $\int dM \langle N_i(M)\rangle dn/dM$,
where $dn/dM$ is the halo mass function and index $i$ is for either central or satellite source occupation number.

\section{Results \& Discussion}

In Fig.~2, we summarize our first set of results related to $w(\theta)$ measurements for each of the three SPIRE bands and for sources 
with $S>30$mJy.
We show correlation functions measured for sources in both Lockman-SWIRE and FLS fields here. We find no statistical difference in the 
correlation functions of sources detected in the two fields down to the flux density cut-off of 30 mJy. 
To test for evolutionary hints in clustering, in Fig.~3 (a) we split the 250 $\mu$m source sample to two bins in flux density, while in Fig.~3(b)
we split the combined sample to two colour bins. 

To model $w(\theta)$  we need to establish the redshift distribution of the source samples.
Given the lack of adequate spectroscopic redshifts, we make use of sub-mm colours to generate a qualitative
redshift distribution (e.g., Hughes et al. 2002).   First we generate $10^6$ isothermal SED models using modified black-body spectra 
with a broad range in dust temperature (10K to 60K) and dust emissivity $\beta$. 
We also include a 15\% Gaussian scatter to model predictions to account for uncertainties in the observed fluxes (Swinyard et al. 2010).
We grid the models along the redshift direction in the colour-colour plane to several bins and simply convert the number  of observed 
data points in each bin in the colour-colour plane to a distribution function in redshift (see Fig.~4 for an example).   
While these distributions are generally consistent with certain model predictions 
(e.g., Le Borgne et al. 2009; Valiante et al. 2009) and with sub-mm galaxy data (e.g., Chapman et al. 2005), 
the redshift distribution we recover is  strongly  sensitive to the SEDs used and should only be considered as an approximate. 

Our model fitting results are summarized in Table~1. We assume WMAP 5-year best-fit $\Lambda$CDM cosmology (Komatsu et al. 2009).
While in Figs.~2 and ~3 we only show the errors from the variances ($\sqrt{C_{ii}}$),  parameter results 
shown in Table~1 account for the covariance matrix when model fitting to measurements (e.g., Fig.~1 where we show the correlation matrix).
Down to the 30 mJy flux density cut, we find 
average  bias factors of $2.9 \pm 0.4$ and $3.6 \pm 0.8$ for 250 and 500 $\mu$m sources, respectively. Fitting a power-law to all data,
the correlation lengths, $r_0$, are $4.5 \pm 0.5$ Mpc (250 $\mu$m) and $6.3 \pm 0.7$ Mpc (500 $\mu$m).
While 250 $\mu$m sources are more likely to be found in halos with mass $(5\pm4)\times10^{12}$ M$_{\sun}$, we find that the bright 500 $\mu$m
sources in our sample occupy halos of $(3.1\pm2.8)\times10^{13}$ M$_{\sun}$. The difference is because at a given redshift the 500 $\mu$m 
sources are at a higher luminosity. 
Our modeling allows us to establish that $(14 \pm 8)$\% of the sources appear as satellites in massive halos than the minimum mass scale. 
In the case of 350 $\mu$m and 500 $\mu$m source samples, we have failed to accurately determine the parameters related to satellite occupation number. 
As a test on the validity of our results to uncertainties in $n(z)$,
we also considered two extreme possibilities by placing all sources either at $z \sim 1.5$ or $z \sim 3$ and
found parameters within 1$\sigma$ uncertainties of the estimates quoted in Table~1. This is mostly due to the fact that
$n(z)$ we use for model fitting , with an example shown in Fig.~4, is broad with a tail to low redshifts.

Our measurements show some evidence for non-linear clustering at arcminute angular scales. 
Compared to 250 $\mu$m, an increase in clustering at arcminute angular scales is less clear at 350 and 500 $\mu$m due to the  increase in the 
beam size.  In comparison, angular power spectra of {\it BLAST} fluctuations did not convincingly reveal a 1-halo term (Viero et al. 2009)
and clustering was found to be even below the linear term at smallest angular scales probed. The increase in arcminute-scale angular clustering we see here
 demonstrates the crucial role played by superior angular resolution of SPIRE.
Beyond this initial study, future work involving understanding the large-scale structure distribution of sub-mm galaxies will pursue 
additional cross-clustering studies of sub-mm sources with optical and shorter IR wavelengths,
and studies of unresolved fluctuations. On the modeling side an approach based on the
conditional luminosity function could be used to extract additional details on the spatial distribution of {\it Herschel} sources.

\begin{acknowledgements}

Cooray, Amblard, Serra, Khostovan, and Mitchell-Wynne are supported by NASA funds for US participants in {\it Herschel} through an award from JPL.
SPIRE has been developed by a consortium of institutes led by Cardiff University (UK) and including Univ. Lethbridge (Canada); NAOC (China); CEA, LAM (France); 
IFSI, Univ. Padua (Italy); IAC (Spain); Stockholm Observatory (Sweden); Imperial College London, RAL, UCL-MSSL, UKATC, Univ. Sussex (UK); and Caltech/JPL, IPAC, 
Univ. Colorado (USA). This development has been supported by national funding agencies: CSA (Canada); NAOC (China); CEA, CNES, CNRS (France); ASI (Italy); MCINN (Spain); 
SNSB (Sweden); STFC (UK); and NASA (USA). The data presented in this paper will be released through the HeDaM Database in Marseille\footnote{hedam.oamp.fr/HerMES}.

\end{acknowledgements}

\end{document}